\documentclass[apjl]{emulateapj}
\shorttitle{Large scale systematic signals in weak lensing surveys}
\shortauthors{Vale et al.}

\begin{document}

\title{Large scale systematic signals in weak lensing surveys}

\author{Chris Vale          \altaffilmark{1},
    Henk Hoekstra           \altaffilmark{2,3}, 
    Ludovic van Waerbeke    \altaffilmark{4}, and 
    Martin White            \altaffilmark{1,5}}

\altaffiltext{1}{Physics Department, University of California, 
  Berkeley, CA 94720}
\altaffiltext{2}{Canadian Institute for Theoretical Astrophysics, 
  University of Toronto, 60 St. George Street, M5S 3H8, Toronto, Canada}
\altaffiltext{3}{Department of Astronomy and Astrophysics, University of
  Toronto, 60 St. George Street, M5S 3H8, Toronto, Canada}
\altaffiltext{4}{Institut d'Astrophysique de Paris, 98 bis, boulevard 
  Arago, F-75014 Paris, France}
\altaffiltext{5}{Astronomy Department, University of California, 
  Berkeley, CA 94720}

\begin{abstract}
We use numerical simulations to model the effect of seeing and
extinction modulations on weak lensing surveys.  We find that systematic 
fluctuations in the shear amplitude and source depth can give
rise to changes in the $E$-mode signal and to varying amplitudes
of large scale $B$-modes.  Exquisite control of such systematics
will be required as we approach the era of precision cosmology with
weak lensing.
\end{abstract}


\section{Introduction}

Weak gravitational lensing of distant galaxies by foreground large scale 
structure has emerged as a powerful tool for modern cosmology 
\citep[see][for reviews]{LensReview1,LensReview2,LensReview3}, which has 
already provided constraints on cosmological parameters 
\citep[see e.g.][for the current status]{HoeYeeGla,WaeMel} and been touted 
for its potential to constrain dark energy
\citep{BenBer,Hut02,Hu02,AbaDod,BenWae03,JaiTay,Hea03,Ref03b,TakJai03,
BerJai,TakWhi}.

The technique relies upon the measurement of the distortion that lensing 
induces in the shapes of galaxy images.  The percent level distortion 
induced by large-scale structure is generally referred to as cosmic shear.  
As the power of cosmic shear surveys increases, requirements that systematic 
errors in the measurement of these images be accurately accounted for has 
become increasingly stringent.  Fortunately nature has provided us with a 
means to test for some of these systematic errors.  Since lensing arises 
from a scalar gravitational potential, the shear pattern it generates has a 
particular form.  For example, the shear pattern around an isolated,
spherically symmetric mass distribution is tangential.  Since this pattern
has even parity it is often referred to as a (positive) $E$-mode.  In the
absence of lens-lens coupling or higher order effects the shear pattern
induced by lensing is pure $E$.  A $45^\circ$ rotation of the shear, to
produce the (parity-odd) $B$-mode, will null such a signal.
Thus, a simple diagnostic test for a wide range of systematic errors is
the presence of a $B$-mode in the lensing maps.

In this paper, we model the effect of systematic errors in the seeing
correction and extinction on simulated weak lensing maps.  We show that
fluctuations in amplitude and depth of the signal across a field can
generate $B$-mode signals on large angular scales.  However these $B$-modes 
do not closely track the change in the $E$-mode power, and thus cannot be 
used to correct for these systematic errors.  

We begin with a brief description of these effects in \S\ref{sec:seeing},
and then describe how we generate simulated weak lensing maps in
\S\ref{sec:simulations}.  We discuss how we model the effects of seeing
and dust extinction in the source population in \S\ref{sec:modeling},
and present our results in \S\ref{sec:results} before concluding in
\S\ref{sec:conclusion}. 

\section{Seeing and Extinction} \label{sec:seeing}

One systematic effect that has received particular attention is the
correction for the point spread function (PSF) \citep[e.g.][]{Kaiser95,
Hoekstra98,Kuijken99,Kaiser00,BerJarv02,HirSel03,Hoekstra04} and its 
anisotropy.  Here, we consider two related issues that might prove troublesome:
the effects of fluctuations in seeing conditions and of galactic extinction.

Seeing causes a degradation in the lensing signal amplitude by circularizing 
the images of background galaxies.  Corrections for this ``isotropic'' PSF 
effect have been studied using simulated images 
\citep[e.g.][]{Bacon01,Erben01} and 
by marginalizing over the uncertain shear calibration using a model fit to the 
power spectrum \citep{IHMS}.  In practice a cosmic shear survey consists of 
many pointings of a telescope, and seeing can easily vary by a factor of two 
between the best and worst pointings;  even chip to chip variations in the 
detector can yield corrective factors that differ by as much as 10\%.  
Systematic errors in the lensing measurement may therefore be introduced on 
the scale of the telescope pointings or of individual chips in the detectors 
due to imperfections in this correction.

A second effect of seeing fluctuations is to modulate the effective depth 
of the survey by down weighting smaller images which have become too 
circularized.  This not only reduces the signal to noise, it also alters 
the source redshift distribution, and can therefore be expected to have an 
effect that resembles source redshift clustering \citep[e.g.][]{Bernardeau98}, 
although on a different angular scale.
The measured signal will then probe structures at different depths in
different regions of the sky.
Variable galactic extinction in a magnitude limited survey will have a
similar effect, although it should not be correlated with the lensing
signal.

\begin{figure*}[!t]
\begin{center}
\resizebox{!}{5.5cm}{\includegraphics{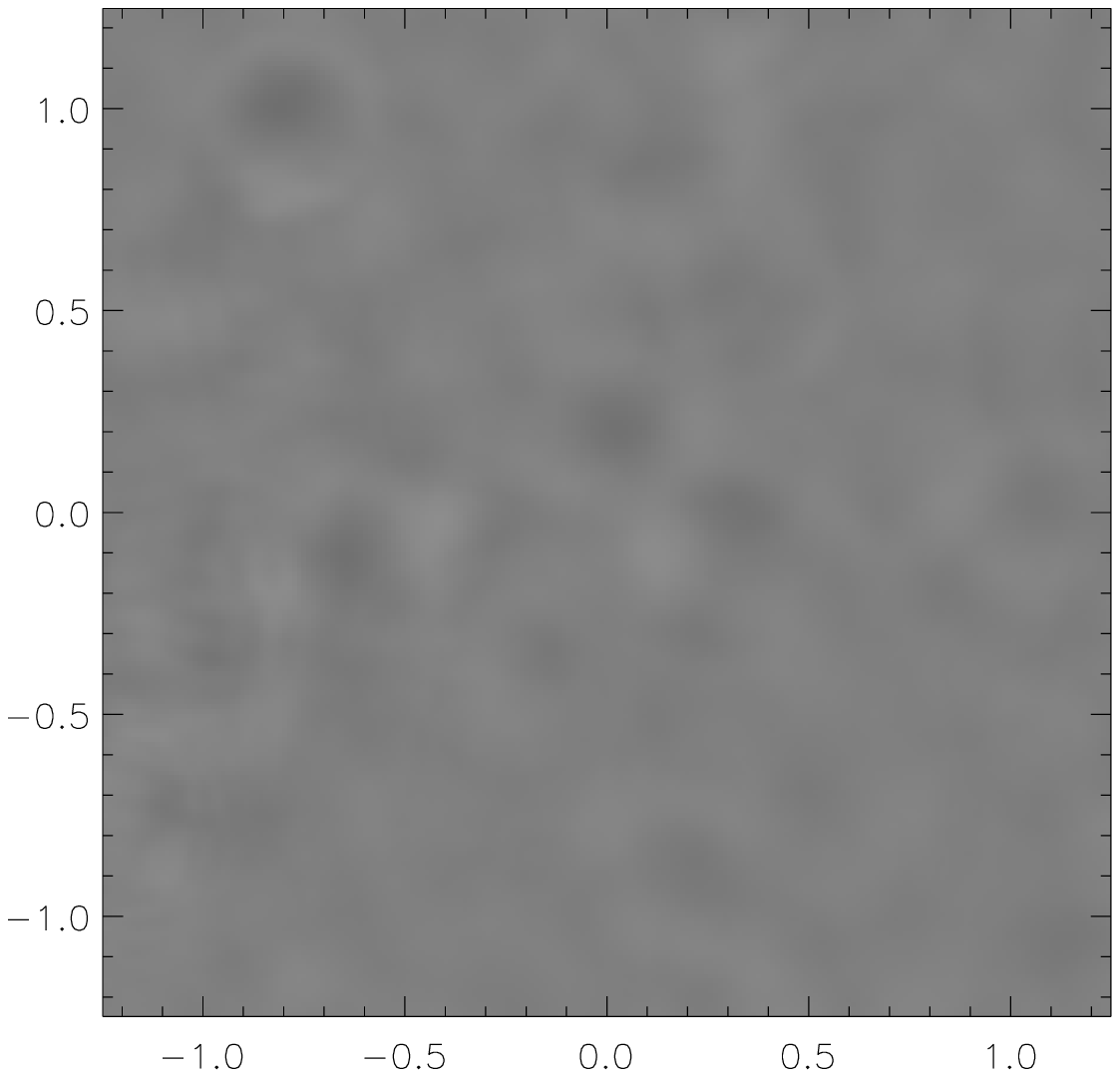}}
\resizebox{!}{5.5cm}{\includegraphics{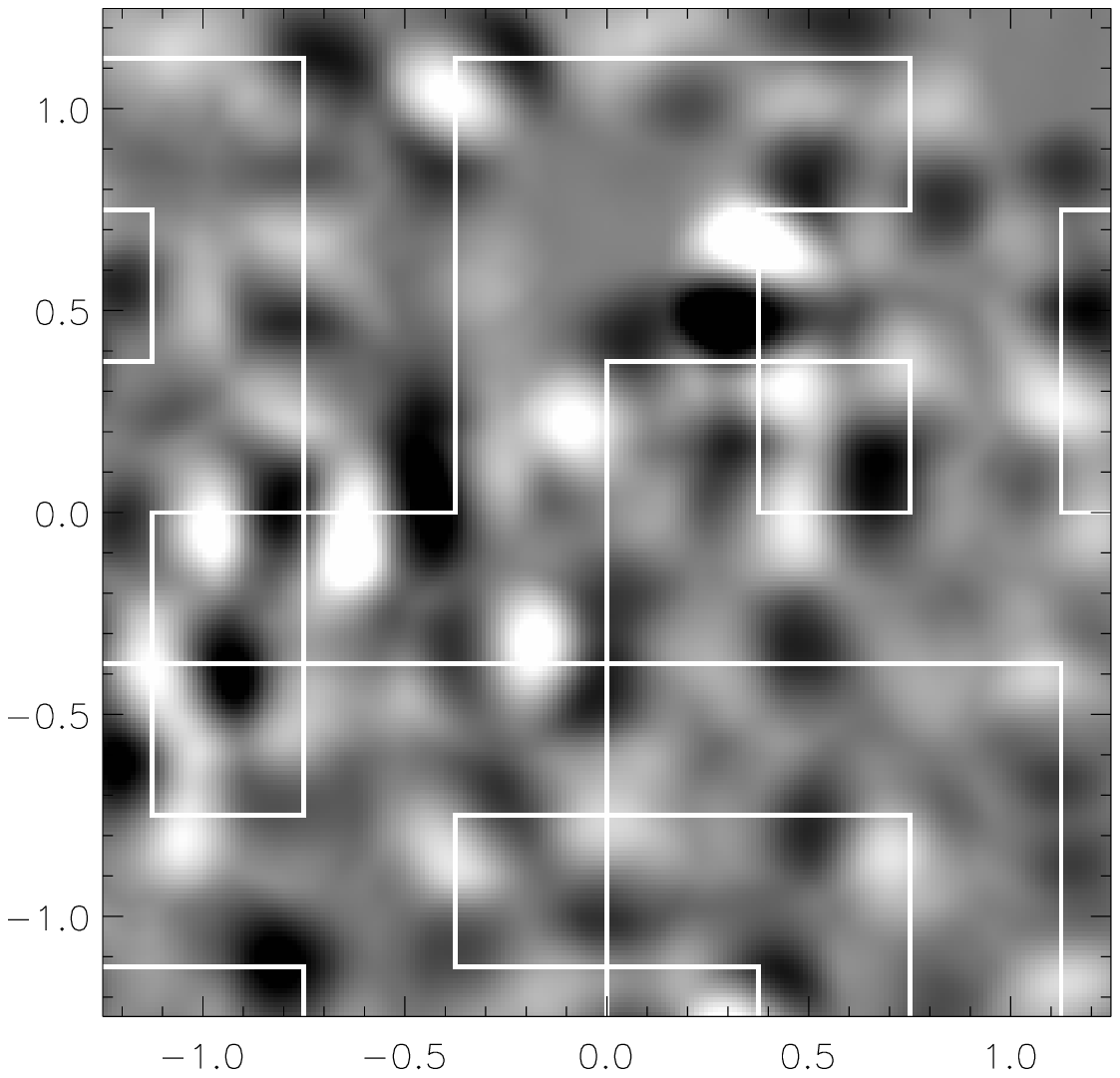}}
\resizebox{!}{5.5cm}{\includegraphics{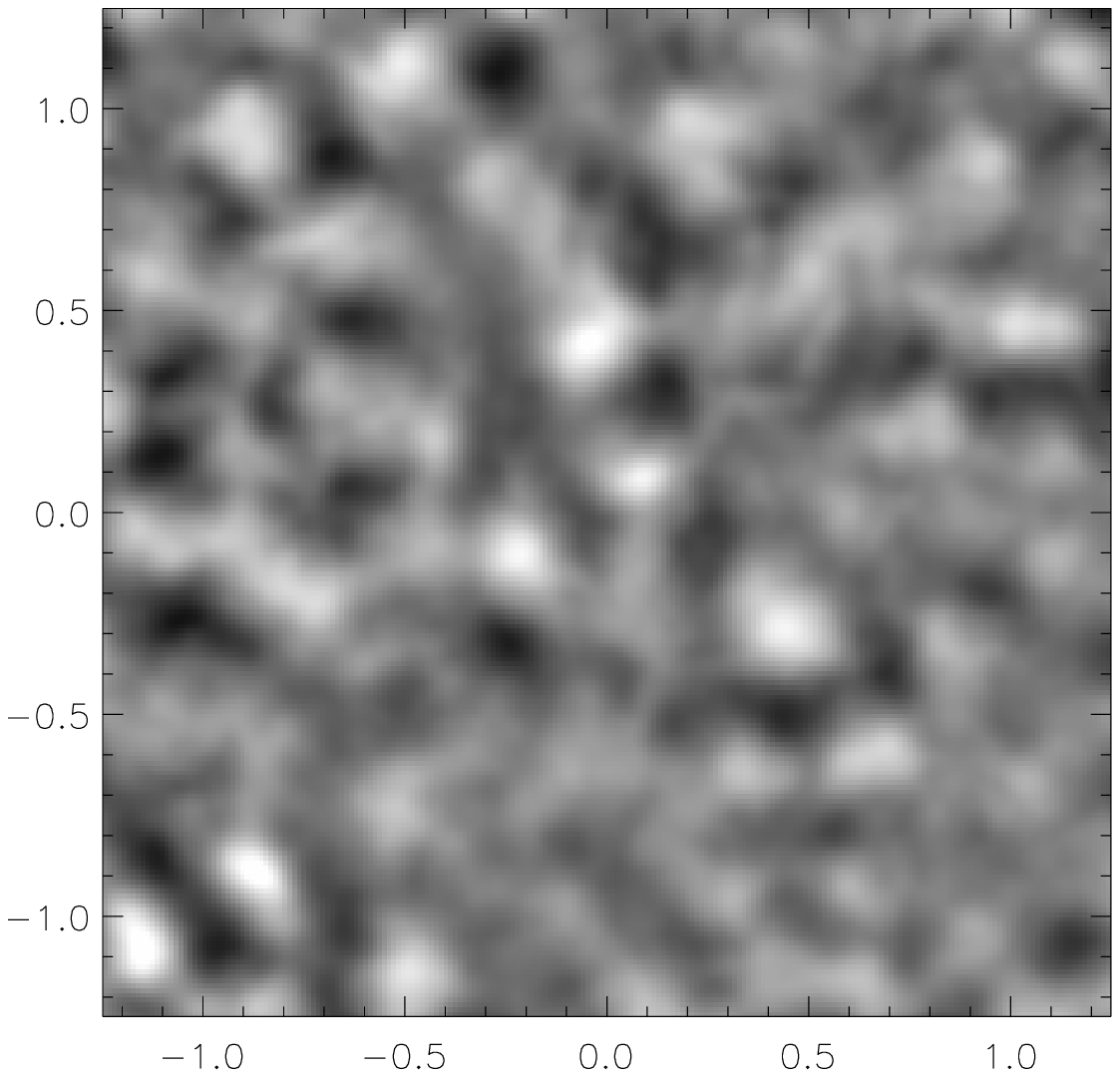}}
\end{center}
\caption{$B$-mode aperture mass on a $15^\prime$ scale, for 
$2.5^\circ \times 2.5^\circ$ fields.  The left panel is with no modulation, 
the center panel includes sharp, 10\%, modulations in the seeing correction
on a $22^\prime$ scale (as detailed in the text and outlined by the solid
lines), and the right panel is for 10\% modulations in the source redshift
relation at a scale of $8^\prime$.
The greyscale is linear and spans $\pm 5 \times 10^{-4}$.}
\label{fig:3panel}
\end{figure*}
\smallskip
\section{Simulations} \label{sec:simulations}

In order to assess the systematic errors introduced by uncorrected changes
in the calibration or depth of the survey we make use of simulated weak
lensing maps.  These maps are generated by ray tracing through N-body
simulations.  We use the methods and models described in detail in 
\cite{WV03}, so we provide only a brief summary here.  

Our calculation is done within the context of a $\Lambda$CDM model
\citep[model 1 of][]{WV03} chosen to provide a good fit to recent CMB and 
large scale structure data.  The weak lensing maps are made from an N-body 
simulation using a multi-plane ray tracing code, as described in \cite{ValWhi}.
The code computes the $2\times 2$ shear matrix $\bf{A}$
\begin{equation}
  A_{ij} = 2 \int { d \chi \ g(\chi) \nabla_i \nabla_j \phi}
\label{eqn:shear}
\end{equation}
where $\bf A$ describes the distortion of an image due to lensing, $\phi$ 
is the gravitational potential, $\chi$ is the comoving distance, and 
$g(\chi)$ is the lensing weight
\begin{equation}
  g(\chi) \equiv \int_\chi^\infty d\chi_s\ p(\chi_s)
    {\chi(\chi_s-\chi)\over\chi_s}
\end{equation}
for sources with distribution $p(\chi_s)$ normalized to $\int dp=1$.  The 
shear matrix $\bf A$ is decomposed as
\begin{equation}
\mathbf{A} = 
  \left( \begin{array}{cc}
  \kappa + \gamma_1     & \ \ \ \gamma_2 + \omega\\
  \gamma_2 - \omega     & \ \ \ \kappa - \gamma_1
  \end{array} \right)
\end{equation}
where the $\gamma_i$ are the shear components, $\kappa$ is the convergence, 
and $\omega$ is the rotation, which is generally small.

We make maps of the shear and the convergence at a range of source redshifts 
from $z\sim 0$ to $3$ in steps of $\Delta \chi = 50\,h^{-1}$Mpc.  
In each case, a $2048^2$ grid of rays subtending a field of view of $3^\circ$ 
is traced through the simulation.  The two shear components and the 
convergence are output at each source plane, and down-sampled to $1024^2$ 
pixels.  The final map is a weighted sum of the contributions from each 
source plane;  for a distribution $dp/dz_s$ the weight given to source 
plane $j$ is
\begin{equation}
  w_j = \left. {dp\over dz_s}\right|_j H(z_j) \Delta\chi  \qquad .
\end{equation}
We use a source distribution of the form \citep{Brainerd}
\begin{equation}
  {dp\over dz_s} \propto z_s^2 \exp\left[ -(z_s/z_0)^{3/2} \right]
\label{eqn:dpdz}
\end{equation}
For this distribution $\langle z\rangle=\Gamma({8\over 3})z_0 \simeq 1.5\,z_0$.
We shall use $z_0 = 2/3$ for our base model, and include fluctuations in
$z_0$ where appropriate.

\section{Modeling Systematics} \label{sec:modeling}

\begin{figure*}[!t]
\begin{center}
\resizebox{3.4in}{!}{\includegraphics{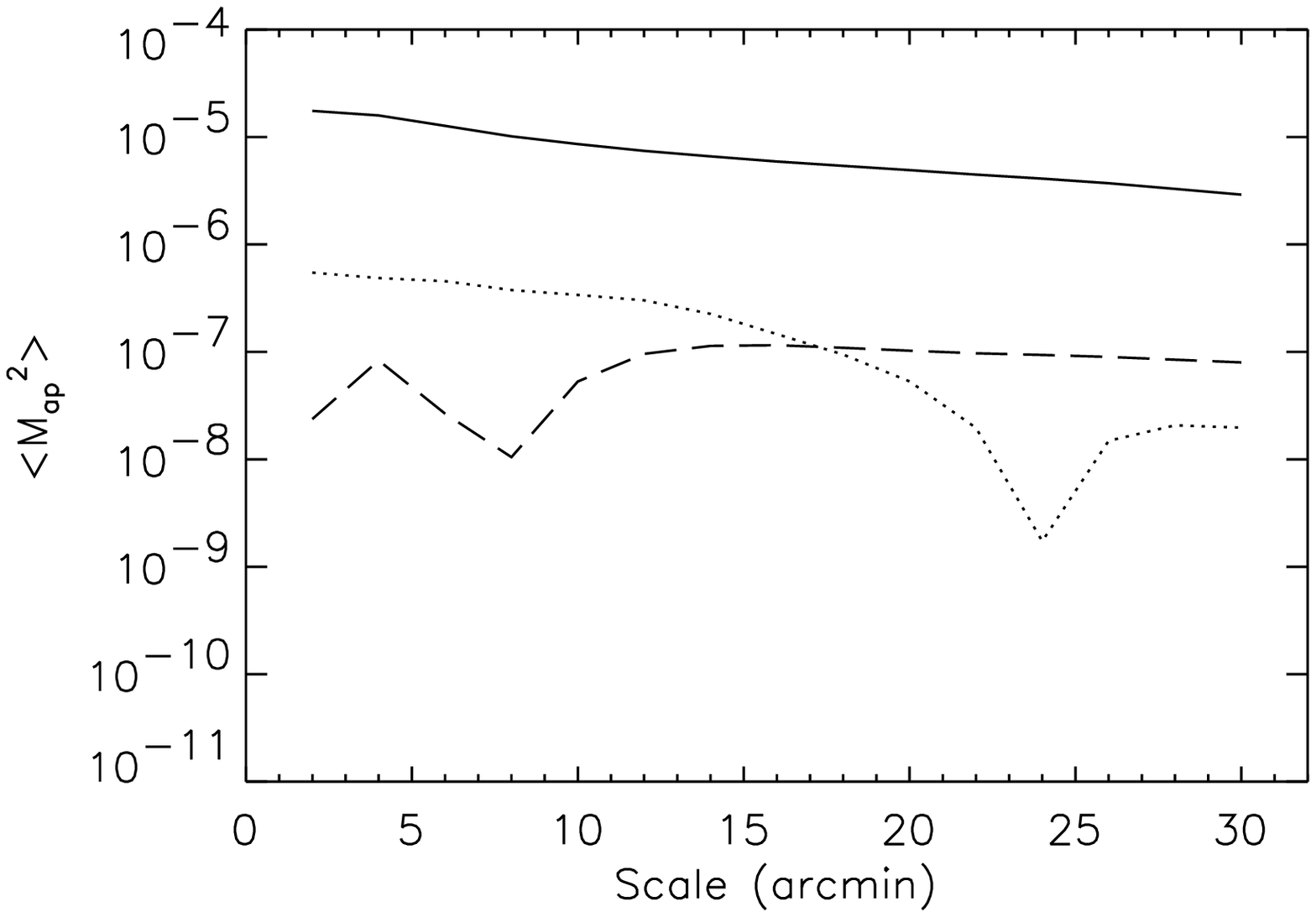}}
\resizebox{3.4in}{!}{\includegraphics{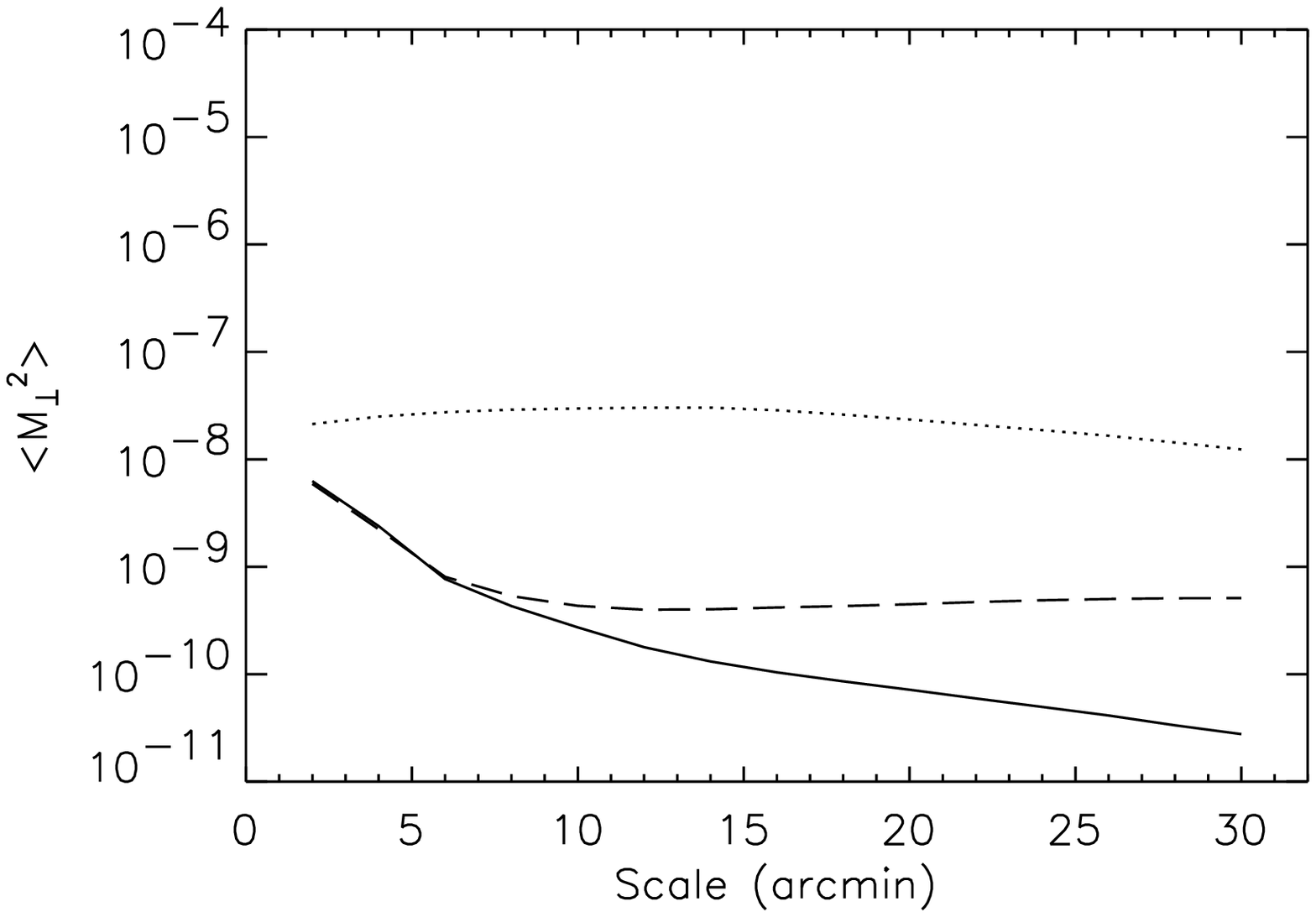}}
\end{center}
\caption{Variance in the aperture mass resulting from 10\% modulations of
the shear amplitude.  (Left) The $E$-mode power (solid), and the absolute
value of the change in the $E$-mode power resulting from modulations using a
pixelized grid (dotted) and a smooth gradient (dashed) as described in
the text.  In both cases the $E$-mode power is reduced on the largest scales 
for this particular realization.
(Right) The $B$-mode power resulting from the same modulations.}
\label{fig:MapAmp}
\end{figure*}

We are interested in the effects of modulations in the calibration of the 
amplitude of the shear signal and in the source redshift distribution.  
The former can arise from variations in the seeing correction for 
different pointings or for different chips, while the latter may occur as 
the result of seeing and extinction variations in a magnitude limited 
survey, as described above. 

We test the first of these effects by generating two ``modulation maps'' of 
unit mean.  The first is a smooth gradient running from 0.9 to 1.1 from one 
side of the map to the other, while the second is generated by dividing the 
map into 64 $22^\prime \times 22^\prime$ regions, each of which is randomly 
assigned a value of either 0.9 or 1.1;  this second modulation map has sharp 
edges, as we expect will occur for chip to chip and pointing variations in the 
seeing correction.  While 10\% fluctuations are toward the upper end of the
range we expect, it serves to illustrate the effects of interest here. 
(A 5\% fluctuation gives rise to $B$-modes a factor of 4 smaller and $E$-modes 
a factor of 2 smaller than a 10\% fluctuation, as expected.)  
These maps are then used to scale the shear at each pixel.

We use a similar tactic to model the source redshift variation, using a 
third ``modulation map'', also of unit mean, but now with a Gaussian 
distribution;  unlike the shear calibration, we do not expect 
the seeing and dust extinction effects to contain sharp edges.  We 
vary the amplitude of the modulation and the coherence scale of the 
fluctuations as follows.  We first generate a Gaussian of zero mean and 
unit variance, with each pixel independent.  We then apply a boxcar 
smoothing to the map with a scale of 4.5, 9 or 18 arcminutes.  The map is 
then scaled to have 5, 10 or 20\% standard deviation and we add 1 to every 
pixel.  The map is clipped so that every pixel lies in the range $[0.1,1.9]$ 
to avoid generating negative or extreme source redshifts.  The resulting map 
is then used to modulate $z_0$ in Eq.~(\ref{eqn:dpdz}) when generating
the shear maps from the sum of the planes described above.  The shear at
each pixel is thus a different weighted sum of the planes.
In each case we find that the effect on different realizations of the
shear field is very similar, so we show results from one
$3^\circ\times 3^\circ$ field below.

\section{Results} \label{sec:results}

To quantify the changes in the $E$- and $B$-modes caused by these modulations
we compute the aperture mass statistics $M_{\rm ap}$ and $M_\perp$ from 
the shear maps.  The former should be sensitive only to $E$-modes, while 
the latter is sensitive only to $B$-modes.  We use the $\ell=1$ form of 
$M_{\rm ap}$ described by \cite{sch98},
\begin{equation}
   M_{\rm ap}(\vec x_0;R) = \int d^2x \ 
   \gamma_{\mbox{$\scriptscriptstyle{T}$}}(\vec x+\vec x_0)
   \ G\left({|\vec x|\over R}\right)
\label{eqn:M_ap}
\end{equation}
where $\vec x_0$ is the position on the sky, $R$ is the angular scale, 
$\gamma_{\scriptscriptstyle{T}} = -(\gamma_1\cos2\phi+\gamma_2\sin2\phi)$
is the tangential shear and
\begin{equation}
  G(y) = {6\over\pi R^2}\ y^2(1-y^2) \ \ \textrm{for} \ y \leq 1
\end{equation}
is the radial kernel, which vanishes for $y>1$.  On the smallest 
scales ($<2'$) the convolution is not well approximated by the sum
over pixels, but this is not an issue for the larger scales which
will be of most interest to us.
To compute $M_\perp$, we interchange $\gamma_1\to\gamma_2$ and
$\gamma_2\to -\gamma_1$ before computing the integral in
Eq.~(\ref{eqn:M_ap}).

\begin{figure*}[!t]
\begin{center}
\resizebox{3.4in}{!}{\includegraphics{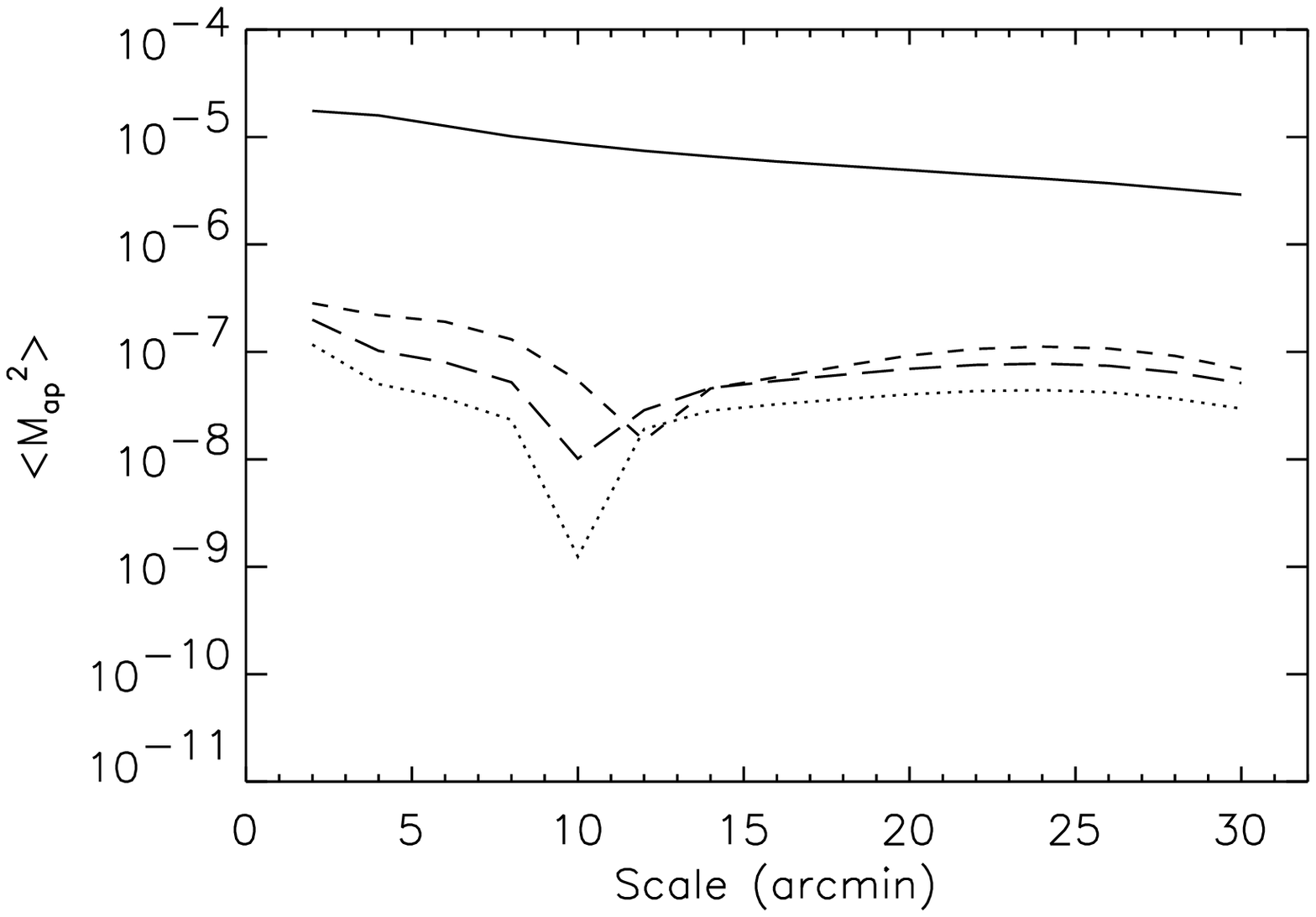}}
\resizebox{3.4in}{!}{\includegraphics{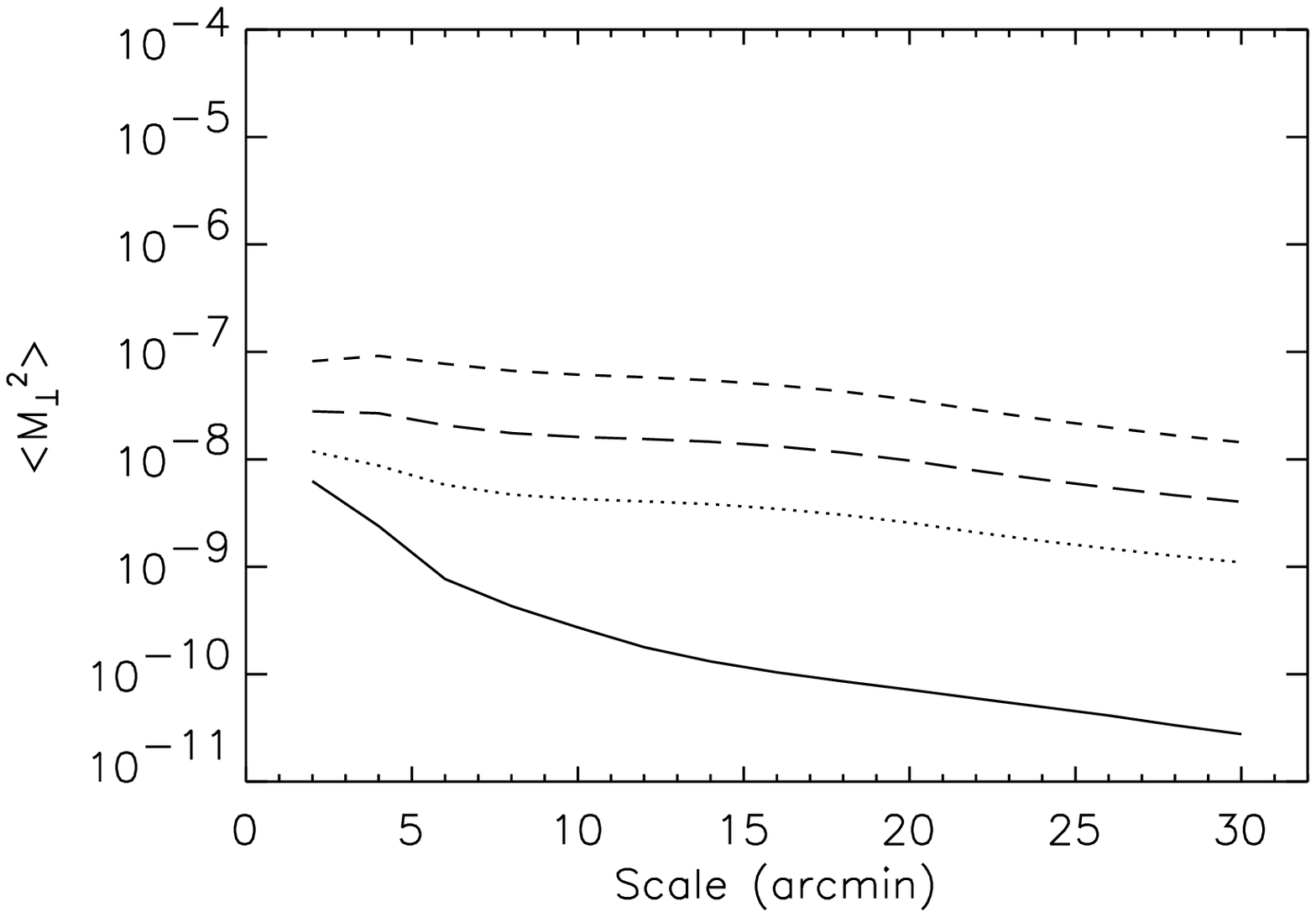}}
\end{center}
\caption{Aperture mass variance from modulations in the mean source redshift,
using a coherence scale in the modulation map of $5^\prime$.
(Left) The unperturbed $E$-mode power (solid) and source redshift modulations
of 5 (dotted), 10 (long dashed) and 20\% (short dashed).
(Right) The $B$-mode power, with line styles as for the $E$-mode.}
\label{fig:MapZ}
\end{figure*}

An example of an $M_\perp$ map, excluding the regions within $R$ of the map 
edges, is given in Figure \ref{fig:3panel}.  Note that even the unmodulated 
maps contain some $B$-mode power, as expected from effects such as lens-lens 
coupling and violations of the Born approximation \citep{JSW,ValWhi}.  
However, modulations of the amplitude of the shear signal and of the source
redshift distribution both significantly enhance the $B$-mode signal.  
While the $B$-modes are concentrated in regions where the amplitude or
depth changes abruptly, there is additional structure in the lensing
signal which makes the pattern somewhat complex.

In order to quantify these effects further, we compute the variance of the 
$M_{\rm ap}$ and $M_\perp$ maps, again excluding the regions within 
$R$ of a map edge.  This variance probes a narrow range of wavemodes 
in the power spectrum, peaked at roughly $1/3$ of the filter scale, $R$.

We show the effect of modulations in the amplitude of the shear in
Figure \ref{fig:MapAmp}.
The smooth, gradient modulation generates very small $B$-modes and a
change in the $E$-mode power on all scales.  This is not unexpected:
the initial $B$-mode is very small.  If the transformation from shear to
convergence ($E$-mode) was completely local then rescaling the shear would
not generate any $B$-mode.  The figure of merit is thus how much the
gradient changes across the ``non-local'' scale in converting shear to
convergence.
Whatever the reason, the $B$-mode is much smaller in amplitude than the
change in the $E$-mode, also shown in Figure \ref{fig:MapAmp}.

A sharp modulation in amplitude gives a larger $B$-mode signal, with a 
similar change in the $E$-mode signal (Figure \ref{fig:MapAmp}).
Below $25'$, the $B$-mode signal in this realization does not track the 
change in the $E$-mode signal, which is decreasing for angles large compared 
to the modulation.  However the amplitudes are more comparable than above, 
becoming very similar at the largest scales probed.

The inclusion of fluctuations in the source redshift distribution increases
the large angle $B$-modes by roughly 2 orders of magnitude, as can be seen
in Figure \ref{fig:MapZ}.
The amplitude of this effect on scales larger than the modulation scale
is (almost) independent of the angular scale of the modulation, but is an
increasing function of the modulation amplitude in all cases.
In extreme cases, the effect can be as large as 1\% of the $E$-mode variance
(i.e. 10\% of the amplitude).
The {\it change\/} in the $E$-mode variance also increases with increasing
modulation, but not as significantly as does the $B$-mode.  

\section{Conclusion} \label{sec:conclusion}

We have used numerical simulations to model the effect of seeing and
extinction modulations on weak lensing surveys.  We find that fluctuations
in both the shear amplitude and the source $z$-distribution can give
rise to changes in the $E$-mode power and to large scale $B$-modes, with
sharp changes in amplitude or fluctuations in the $z$-distribution giving 
the larger $B$-modes.
Since the $B$-modes do not closely track the changes in $E$-mode power
they cannot be used to ``correct'' for the above effects.
In the case of strong $B$-mode enhancement, however, they can be used as
a monitor for the effect.

As we move from first detections into scientific exploitation of cosmic
shear, effects such as this will need to be carefully controlled.  
Photometric redshift information offers a likely route to mitigating 
fluctuations in the survey depth while offering many scientific 
advantages.  The stable and excellent observing conditions from space can 
be expected to largely eliminate effects from pointing and seeing 
fluctuations.  Regardless of the route, it is clear that systematic errors 
such as those we have discussed here must be controlled if we are to 
realize the full power of upcoming weak lensing surveys, which will help 
usher in a new era in precision cosmology.

\smallskip
The simulations used here were performed on the IBM-SP at the National
Energy Research Scientific Computing Center.
This research was supported by the NSF and NASA.

\end{document}